\documentclass{emulateapj}
\shorttitle{Jet-Shocked Molecular Gas in NGC 4258}

\begin{document} 
\title{Jet-Shocked H$_2$ and CO in the Anomalous Arms of Molecular Hydrogen Emission Galaxy NGC 4258}
\author{Ogle, P. M., Lanz, L., Appleton, P. N.}
\affil{Infrared Processing and Analysis Center, California Institute of Technology}
\affil{1200 E. California Boulevard, Pasadena, CA, 91125}
\email{ogle@ipac.caltech.edu}

\begin{abstract}

We present a {\it Spitzer} Infrared Spectrograph (IRS) map of H$_2$ emission from the nearby galaxy NGC 4258 (Messier 106). The H$_2$ emission
comes from $9.4 \pm 0.4 \times 10^6 M_\odot$ of warm molecular hydrogen heated to 240-1040 K in the inner anomalous arms, a signature of jet interaction
with the galaxy disk. The spectrum is that of a molecular hydrogen emission galaxy (MOHEG), with a large ratio of H$_2$ over 7.7$\mu$m PAH emission (0.37),
characteristic of shocked molecular gas.  We find close spatial correspondence between the H$_2$ and CO emission from the anomalous arms. Our estimate of
cold molecular gas mass based on CO emission is 10 times greater than our estimate of $1.0\times 10^8 M_\odot$ based on dust emission. We suggest that the
$X_{CO}$ value is 10 times lower than the Milky Way value because of high kinetic temperature and enhanced turbulence. The H$_2$ disk has been overrun and
is being shocked by the jet cocoon, and much of the gas originally in the disk has been ejected into the galaxy halo in an X-ray-hot outflow. We measure a
modest star formation rate of 0.08 $M_\odot$ yr$^{-1}$ in the central 3.4 kpc$^2$ that is consistent with the remaining gas surface density.
  
\end{abstract}


\section{Introduction}

\subsection{Radio Jet Feedback}

Radio jet-interstellar medium (ISM) interactions may cause either positive or negative feedback on star formation in galaxies \citep{wb11}.
Jet feedback may explain the truncation of the galaxy mass function and quenching of star formation via expulsion or heating of the ISM \citep{csw06}.
At larger scales, radio jets may prevent the cooling and accretion of intergalactic medium \citep{fse00}. Hydrodynamical simulations of radio jets
demonstrate that the jet deposits a large amount of energy, inflating a cocoon of hot gas that may encompass and shock the entire host galaxy ISM \citep{sb07}.
Molecular gas may be dissociated by shocks and driven out of the galaxy by outflows, preventing star formation altogether. Alternatively, jet driven turbulence
may support the gas against collapse, preventing star formation.

\subsection{MOHEG Phenomenon}

\cite{obg10, oaa07} find that 30\% of 3C radio galaxies have extremely luminous H$_2$ emission from large
masses (up to $10^{10} M_\odot$) of warm molecular gas. The H$_2$ is heated to temperatures of 100-1500 K.
UV excitation in photodissociation regions (PDRs) is ruled out by a very high ratio of L(H$_2$ 0-0 S(0)-S(3))/L(PAH 7.7 $\mu$m)$>0.04$.
Heating by X-rays from the AGN is ruled out by insufficient X-ray luminosity. The remaining viable sources of heating
are shocks driven into the ISM by the radio jet, or cosmic rays \citep{obg10}. 

Similar conditions and IR spectra are seen in the Stephan's Quintet (SQ) intergalactic shock, where there is a close
correspondence between X-ray, H$\alpha$, H$_2$, [C II], and CO emission \citep{a06,cab10,gbp12, a13}. The large range of densities,
temperatures, and cooling timescales are characteristic of a shocked multiphase medium, where X-ray emitting gas occupies the
lower-density spaces in the post-shocked gas \citep{gbp09}.

X-ray emission from jet-shocked gas is seen in several radio MOHEGs, including Per A \citep{fst06}, Cen A \citep{knb07},
and 3C 305 \citep{mcg09,hmh12}. Some radio MOHEGs display jet-driven neutral and ionized outflows
with velocities up to 1200 km s$^{-1}$ \citep{mto05,emt05,goe12}. The limited spatial resolution of mid-IR H$_2$ observations
of distant radio galaxies has made it difficult to determine the relationship between the radio jet, X-ray emitting gas, outflows,
and the H$_2$ emission region. It is unknown whether the H$_2$ emission comes from a distinct shock front or is distributed
more widely in the galaxy. It is also unclear whether the H$_2$ emission arises from preexisting molecular gas
that has been shocked in-situ, entrained in an outflow, or created in the postshock gas. 

\subsection{Anomalous Arms  and Jet Feedback in NGC 4258}

NGC 4258 (Messier 106, $z=0.001494$, $d=7.2$ Mpc)  is a nearby galaxy hosting a Seyfert 1.9 AGN \citep{h97}. It is known for
its anomalous spiral arms, which emerge from the galaxy nucleus and appear to intersect the regular spiral arms of the galaxy \citep{cc61}. It is
famous for water maser emission from a nuclear disk that enables accurate measurement of the central black hole mass \citep{mmh95,gjm95}.

The anomalous spiral arms of NGC 4258 emit strongly in radio continuum \citep{vom72}, optical nebular lines, and X-rays. Optical emission line ratios
\citep{cmv95} and X-ray spectroscopy \citep{cwd95,wyc01,ylw07} indicate  $10^5-10^7$ K shocked gas in the anomalous arms. Large lobes of X-ray
emitting plasma extend above and below the galactic disk, from a jet-driven outflow \citep{cwd95,wyc01}. A series of H$\alpha$-emitting
streamers of ionized gas branch off of the anomalous arms, marking cooler gas entrained by the hot outflow.

The 3-dimensional orientation of the inner radio jet relative to the maser disk is well-constrained by VLBI observations. Jet hot spots
indicate a consistent jet orientation at larger scales, at an angle of 30\arcdeg~ to the plane of the galaxy disk \citep{cdg00}.
\cite{wyc01} propose that the anomalous arms are the projection of the jet along the disk, along a ``line of destruction''.
Gas is blown out the opposite side of the disk to the jet. 

\section{Observations and Results}

A map of NGC 4258 was made with {\it Spitzer} IRS SL1,2 and LL1,2 modules (PID 30471, PI J. D. Smith). The 60 s SL observations were stepped perpendicular
to the slit $29 \times 1.85\arcsec$. The 30 s LL observations were stepped perpendicular to the slit $11 \times 5.08\arcsec$ and parallel to the
slit $4 \times 15\arcsec$. The inner $50.7\arcsec \times 55.0\arcsec$ ($1.76 \times 1.92$ kpc) of the galaxy is covered by all 4 modules. There is extended coverage
along the SL and LL slits, which are oriented in perpendicular directions. Spectral cubes were constructed from each IRS module with the IDL program
CUBISM \citep{sad07a}, using the off-pointed slit for background subtraction, and 5$\sigma$ global bad pixels were cleaned from the data. 

The integrated IRS spectrum is shown in Figure 1a. The spectrum was extracted from the spectral cubes in the overlap region between the SL and LL slits
indicated in Figure 2. Line and polycyclic aromatic hydrocarbon (PAH) feature fluxes (Table 1) were fit using PAHFIT \citep{sdd07b}. The H$_2$ pure
rotational lines are extraordinarly strong relative to the PAH and continuum emission. Ionized emission lines include [Ar {\sc ii}] $\lambda$6.99,
[S {\sc iv}] $\lambda$10.51, [Ne {\sc ii}] $\lambda$12.81, [Ne {\sc iii}] $\lambda$15.56, [S {\sc iii}] $\lambda$18.71, [O {\sc iv}] $\lambda$25.89,
[Fe {\sc ii}] $\lambda$26.00, [S {\sc iii}] $\lambda$33.48, and [Si {\sc ii}] $\lambda$34.82. The high ionization lines of [Ne {\sc vi}] $\lambda$7.65,
[Ne {\sc v}] $\lambda$14.3, and [Ne {\sc v}] $\lambda$24.31 are undetected.

We extracted slices from the SL1 and LL2 cubes at the wavelengths of the H$_2$ 9.66 $\mu$m 0-0 S(3) and 17.03 $\mu$m 0-0 S(1) lines.
Ancillary images were collected from GALEX (FUV, NUV), SDSS DR10 (u, g, r, i, z), 2MASS (J, H, K), {\it Spitzer} IRAC (3.6, 4.5, 5.8, 8.0$\mu$m),
MIPS 24 $\mu$m, {\it Herschel} PACS (70, 100, 160 $\mu$m) and SPIRE (250, 350 $\mu$m). We extracted matched photometry in each band, inside the IRS
spectral extraction region, including the AGN, and used these to form the spectral energy distribution (SED, Figure 1b). We find an upper limit
of 40\% on the percentage AGN contribution to the PACS photometry within a $12\arcsec$ radius.

The IRS H$_2$ maps in Figure 2 are overlaid on the {\it Spitzer} IRAC 8 $\mu$m image, with CO 1-0 emission contours mapped by BIMA
\citep{htr03}. A primary ridge of H$_2$ emission is located 6\arcsec~ SW of the AGN, running SE-NW along the edge of the anomalous arms. The brightest H$_2$ S(3)
emission appears along the inner edge of this ridge, where it is closest to the AGN and jet. The ridge bends to the NE at a location
20\arcsec~ NW of the AGN. A secondary ridge of emission is located 6\arcsec~ NE of the AGN. The H$_2$ ridges correspond closely to the ridges of CO emission
\citep{mrn89,kfn07}. There is no clear relationship between H$_2$ emission and the PAH and hot dust emission in the IRAC 8 $\mu$m image, consistent with the
results of \cite{lkt10}, who find no obvious impact of the jet on the distribution of 8 $\mu$m emission.

We compare our {\it Spitzer} H$_2$ map with a {\it Chandra} 0.3-8 keV image of the anomalous arms, retrieved from the archive (combined ObsIDs 350, 1618, 2340;
PI A. Wilson). The {\it Chandra} image has been cleaned of extranuclear point sources and smoothed. The clear association between H$_2$ and X-ray emission from the
inner anomalous arms (Fig. 3) suggests that both are powered by jet-driven shocks. The X-ray emitting gas is located either in gaps between the molecular
clouds, or in streams of ejecta directed out of the plane of the galaxy. The H$_2$ emitting gas appears to be confined to the plane, and does not follow the anomalous
arms out of the plane. However, the H$_2$ emission does extend well beyond the projection of the radio jet axis. We suggest that the warm H$_2$ and hot,
intercloud X-ray emitting gas are in the turbulent post-shocked layer left behind as the jet cocoon has expanded and overrun a multi-phase medium.

The inner X-ray jet lies in a channel between two molecular ridges (Fig. 4). An anticorrelation between the H$_2$ and X-ray emission is clearly seen at 6-9\arcsec~
by comparing their azimuthal profiles (Fig. 4). A similar anticorrelation is seen between H$\alpha$ emission and CO emission at these distances \citep{mrn89}. The jet
may have carved out a hole in the molecular gas disk, or alternatively the X-ray and H$\alpha$ emitting gas is confined by a preexisting hole in the disk.

\section{Discussion}

\subsection{H$_2$ Temperature Distribution and Heating}

We created an excitation diagram for the H$_2$ pure rotational lines, and used this to estimate the mass of warm H$_2$ as a function of temperature, assuming
thermal equilibrium. Three temperature components are required to fit the excitation diagram, with
$M$(H$_2$) $=8.9 \pm 0.3 \times 10^6 M_\odot$, $5.3\pm 2.4 \times 10^5 M_\odot$, and $2.5\pm 0.8 \times 10^4 M_\odot$ at $T=240$ K, 450 K, and 1040 K, respectively.
The total warm H$_2$ mass at $T>200$ K is therefore $M_\mathrm{tot}$(H$_2$) $=9.4 \pm 0.4 \times 10^6 M_\odot$.

The mid-IR (MIR) line luminosities inside our IRS extraction region are given in Table 1. The total ionized gas MIR line luminosity is L(MIR,ionized) $= 2.27 \times 10^{40}$ erg s$^{-1}$,
compared to the total H$_2$ 0-0 S(0)-S(7) rotational line luminosity of  L(H$_2$ 0-7) $= 5.27 \times 10^{40}$ erg s$^{-1}$. H$_2$ emission is one of the
main ISM cooling channels. The luminosity of H$_2$, summed over the first 4 transitions is L(H$_2$) $= 3.92 \times 10^{40}$ erg s$^{-1}$, while the PAH 7.7$\mu$m
luminosity is L(PAH 7.7) $= 1.07 \times 10^{41}$ erg s$^{-1}$. The ratio is $L$(H$_2$)/$L$(PAH 7.7) $=0.37$, which qualifies NGC 4258 as a MOHEG \citep{obg10}. A large ratio of H$_2$ to
PAH 7.7 $\mu$m ($>0.04$) indicates that the H$_2$ is not heated by UV photons from hot stars in a PDR \citep{goe12}.

We measure an unabsorbed 2-10 keV luminosity of $7.3 \times 10^{40}$ erg s$^{-1}$ for the AGN from the {\it Chandra} image. The ratio of
L(H$_2$)/$L_\mathrm{X}$(AGN, 2-10 keV) = 0.54 rules out AGN heating of the H$_2$ by X-rays from the nucleus, which has a maximum ratio of 0.01 \citep{obg10}.
The extended thermal X-ray emission inside the IRS extraction region has a total 0.3-2 keV luminosity of  $7.3 \times 10^{39}$ erg s$^{-1}$, which is a factor of
$\sim 7$ less than the total H$_2$ line luminosity, and therefore contributes a negligible amout of X-ray heating to the H$_2$.

Heating by mechanical energy from the jet via shocks or cosmic rays is the prime candidate for powering the H$_2$ emission.
The radio jet power dissipated at its terminal hotspots is estimated to be $L_\mathrm{mech} \simeq 1.9\times 10^{40}$ erg s$^{-1}$, based
on the H$\alpha$ luminosities of the N and S bow shocks, assuming an 0.82\% conversion efficiency \citep{cdg00}. This is only 36\% of the
total H$_2$ power measured within the IRS spectral extraction region. In order to power the observed H$_2$ emission, the jet must be at least
three times more powerful than the above estimate.  Additional power may be dissipated along the entire length of the jet, particularly where
the primary H$_2$ emission ridge is brightest, near the AGN (Figures 2 and 3).

\subsection{CO and Dust Emission}

We measure the BIMA CO 1-0 flux \citep{htr03} in our IRS extraction region to be 1783 Jy km/s. Using the standard Galactic conversion factor of
$X_\mathrm{CO}= 2 \times 10^{20}$ cm$^{-2}$/(K km s$^{-1}$) \citep{l83,bwl13}, we estimate a cold H$_2$ mass of $9.7\times10^{8} M_\odot$.
This is in agreement with previous CO measurements and H$_2$ mass estimates of $\sim 1\times10^{9} M_\odot$\citep{cd96,kfn07}. However, we find evidence
for a nonstandard $X_{CO}$ value below, so this is likely an overestimate of the molecular gas mass.

The integrated 42-122 $\mu$m luminosity from the IRS extraction region is $L$(FIR)$=2.2\times 10^{42}$ erg s$^{-1}$. 
We estimate a dust mass of $M_\mathrm{dust}=(0.8-1.7)\times10^{6} M_\odot$ by fitting the FIR SED with a modified
blackbody with dust emissivity powerlaw index $\beta=1.5-2.0$; in agreement with a dust mass of
$M_\mathrm{dust}=1.0^{+0.2}_{-0.1} \times 10^6 M_\odot$ derived by fitting the UV-FIR SED (Fig. 1b) using {\sc MAGPHYS} \citep{dce08}.
We use this to estimate a total gas mass of $1.0\times 10^8 M_\odot$, assuming a gas/dust mass ratio of 100, as found in the Milky Way \citep{sm79}.
This is only 10\% of the molecular gas mass estimated from CO using the standard $X_{CO}$ value. The corresponding gas surface density is
$\Sigma_\mathrm{gas}=30 M_\odot$ pc$^{-2}$.

To reconcile the large molecular gas mass estimated from CO with the smaller total gas mass estimated from FIR dust emission, either the
dust/gas ratio must be a factor of 10 smaller, or the $X_{CO}$ value must be a factor of 10 lower than in the Milky Way.  While we can not
rule out a low dust/gas ratio in the H$_2$ emission region, dust absorption is seen in optical images of this region with a morphology 
similar to the H$_2$ emission.  In support of a low $X_{CO}$ factor is the close spatial correspondence between CO 1-0 emission and
H$_2$ emission. If most CO emission comes from the same 240-1040 K gas as the H$_2$ emission, we expect the CO emission to be
enhanced by the high temperature and any additional turbulence. Some other nearby galaxies show similarly depressed $X_{CO}$ values in their
central regions \citep{slw13}. The warm H$_2$ in NGC 4258 constitutes roughly 1\% of the CO-derived molecular mass, or 10\% of the dust-derived
gas mass. This warm H$_2$ percentage is similar to that seen for LINER and Seyfert nuclei in the SINGS sample \citep{rhh07}.

\subsection{Jet Feedback on ISM and Star Formation}

The radio jet has a clear impact on the temperature and spatial distribution of molecular and atomic gas in the anomalous arms
of NGC 4258. The question remains whether it has a significant impact on the gas content, star formation rate, and evolution of the
central regions of NGC 4258 and the galaxy as a whole. The clearest indication that the jet is ejecting large quantities gas from the
galaxy is seen in X-ray images, where bubbles of hot gas extend far above and below the plane.  Integrating over the entire X-ray lobes
and assuming a filling factor of $f \lesssim 1$, we find $1.8\times 10^{8} M_\odot f^{0.5}$ of 0.2-0.9 keV gas cooling at a rate of $0.31 M_\odot$ yr$^{-1}$.
An outflow of hot gas at this rate is necessary to sustain the X-ray emission from the lobes, assuming no additional source of heating. If
the gas in the center of the galaxy is the source of this outflow, then most of it has already been ejected into the galaxy halo and the rest
would be depleted in $3\times 10^8$ yr, if not replenished.

We estimate the star formation rate in the IRS extraction region using two methods. The {\sc MAGPHYS} model (Fig. 1b) gives a best-fit
star formation rate (SFR) of $0.084^{+0.015}_{-0.004}$ $M_\odot$ yr$^{-1}$ and a stellar mass of $7.6^{+0.3}_{-1.9}\times 10^9 M_\odot$. In agreement with this, we
estimate SFR(PAH 7.7) $=2.4\times 10^{-9}L$(PAH 7.7)$=0.069$ $M_\odot$ yr$^{-1}$ from the luminosity of the 7.7$\mu$m PAH feature \citep{rsv01}. In comparison,
the Kennicutt-Schmidt (KS) law \citep{k98} predicts a SFR surface density of $\sim 0.03$ $M_\odot$ yr$^{-1}$ kpc$^{-2}$ for the dust-derived gas surface density of
$\Sigma$(H$_2$)$=30 M_\odot$ pc$^{-2}$, or a total SFR of $\sim 0.1$  $M_\odot$ yr$^{-1}$, in agreement with the observed SFR. However, if were to use the
CO-based molecular gas mass density of $\Sigma$(H$_2$)$=300 M_\odot$ pc$^{-2}$, the KS law would predict a 25 times greater SFR of 2.5 $M_\odot$ yr$^{-1}$,
in disagreement with the observed vaule. This demonstrates the crucial role the $X_{CO}$ factor plays in estimates of SFR suppression, which may be greatly
overestimated in MOHEGs where CO emissivity is enhanced and the $X_\mathrm{CO}$ value is low.

\section{Conclusions}

We find H$_2$ emission aligned with the inner anomalous arms of NGC 4258, tracing a large mass of shocked molecular gas. The high H$_2$/PAH 7.7 $\mu$m
ratio of 0.37 is similar to radio galaxies and other MOHEG systems. The H$_2$ emission appears to be powered by the jet-ISM interaction, which drives
shocks into the galaxy disk and generates the outflow of hot gas seen in X-ray emission. However, the jet must be at least 3 times more powerful than
previous estimates in order to drive the observed H$_2$ luminosity of L(H$_2$) $= 5.27 \times 10^{40}$ erg s$^{-1}$. The standard $X_\mathrm{CO}$ overpredicts
the molecular gas mass by a factor of 10 compared to the FIR-derived gas mass of $1.0\times 10^8 M_\odot$. It is likely that CO emission is enhanced and
$X_\mathrm{CO}$ is suppressed by the high temperature and turbulent motions in the jet-shocked anomalous arms. Taking this into account, and using the lower,
FIR-derived gas mass, there is no indication of significant suppression of the SFR in the central regions of NGC 4258 relative to normal galaxies.  While the
SFR is not suppressed in the remaining central molecular disk of NGC 4258, we estimate that most of the gas originally there has already been ejected into the
X-ray emitting lobes, and that the rest will be ejected in $3\times 10^8$ yr.

\acknowledgements

This work is based in part on observations made with the Spitzer Space Telescope, data from the NASA/IPAC Extragalactic Database (NED), and data from the
NASA/ IPAC Infrared Science Archive (IRSA), which are operated by the Jet Propulsion Laboratory, California Institute of Technology, under contract with NASA.
This work is based in part on observations made with Herschel, a European Space Agency Cornerstone Mission with significant participation by NASA.
The scientific results reported in this article are based in part on data obtained from the {\it Chandra} Data Archive and published previously in
cited articles.

\begin{deluxetable}{llllllllll}
\tabletypesize{\small}
\tablecaption{Emission Lines and PAH Features}
\tablehead{ &          &          &         &         &         &         &         &        &         \\}

\startdata
Line         & H$_2$ S(7) & H$_2$ S(6)  & H$_2$ S(5) & H$_2$ S(4) & H$_2$ S(3) & H$_2$ S(2) & H$_2$ S(1) & H$_2$ S(0)& PAH 7.7 \\ 
$\lambda$($\mu$m)    & 5.55  & 6.11  & 6.91  & 8.03  & 9.66  & 12.28  & 17.03  & 28.22  &  7.7  \\ 
Flux         & 4.1   & 1.2   &10.1   & 4.70  &20.47  & 10.45  & 29.49  &  2.8   &172.   \\ 
Flux Unc.    & 0.4   & 0.2   & 0.3   & 0.05  & 0.03  &  0.02  &  0.04  &  0.4   &  1.   \\ 
Luminosity   & 0.26  & 0.08  & 0.64  & 0.30  & 1.29  &  0.66  &  1.86  &  0.18  & 10.8  \\ 
             &       &       &       &       &       &        &        &        &       \\ 
Line         &[Ar {\sc ii}]  & [S {\sc iv}] & [Ne {\sc ii}]  & [Ne {\sc iii}] & [S {\sc iii}]  & [O {\sc iv}]   & [Fe {\sc ii}] & [S {\sc iii}]  & [Si {\sc ii}] \\  
$\lambda$($\mu$m)    & 6.99  & 10.51  & 12.81  & 15.56  & 18.71  & 25.89  & 26.00  & 33.48  & 34.82  \\  
Flux         & 3.5   &  0.94  &  5.28  &  3.66  &  2.98  &  1.3   &  1.5   &  4.796 & 13.179 \\  
Flux Unc.    & 0.3   &  0.02  &  0.02  &  0.03  &  0.03  &  0.5   &  0.4   &  0.002 &  0.003 \\
Luminosity   & 0.22  &  0.06  &  0.33  &  0.23  &  0.19  &  0.08  &  0.09  &  0.30  &  0.83  \\
\enddata
\tablecomments{{\it Spitzer} IRS line flux ($10^{-13}$ erg s$^{-1}$ cm$^{-2}$ Hz$^{-1}$) and luminosity ($1\times10^{40}$ erg s$^{-1}$) within central
                $1.76 \times 1.92$ kpc of NGC 4258.} 
\end{deluxetable}

\begin{figure*}[t]
\centerline{\includegraphics[width=0.8\linewidth]{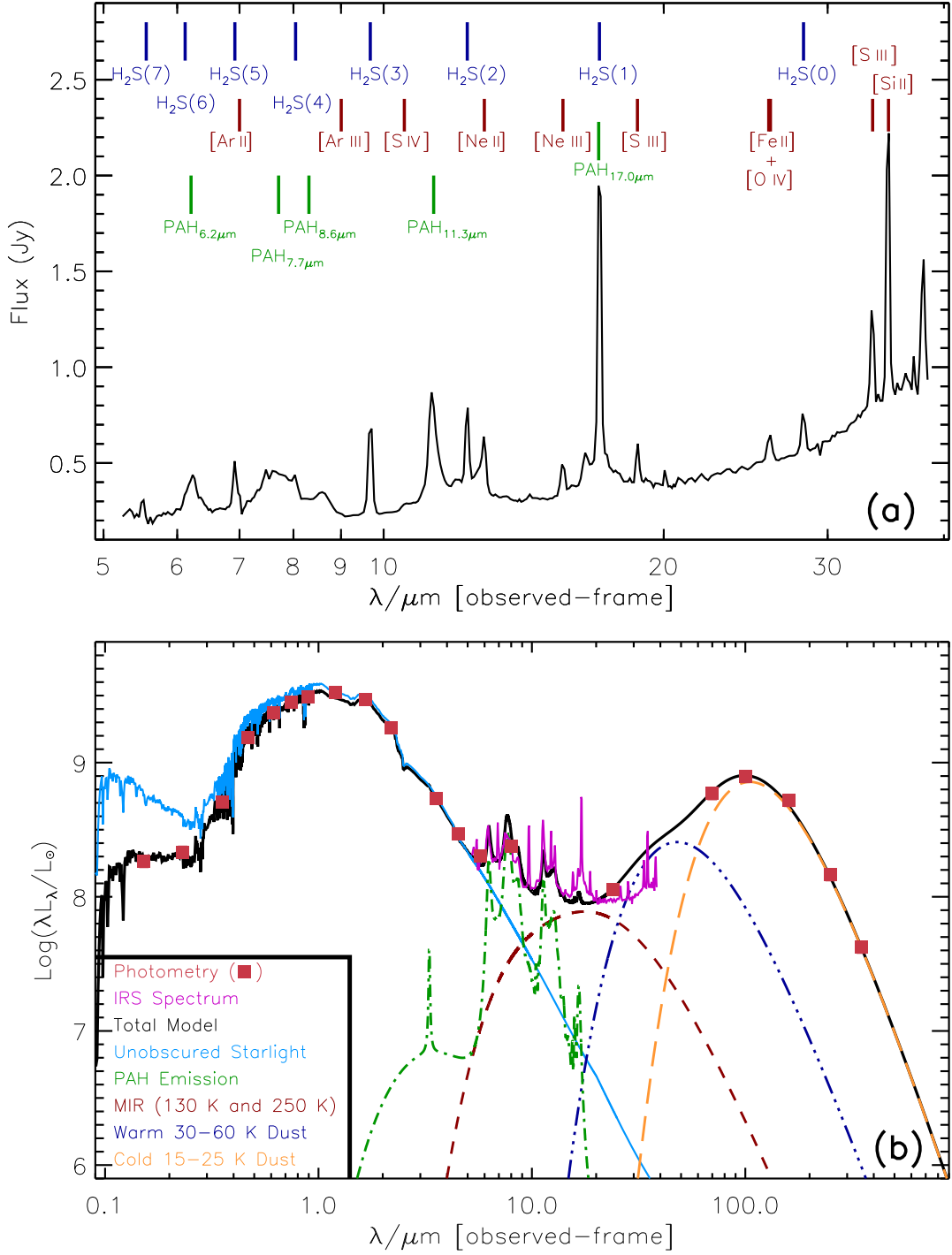}}
\caption{(a) {\it Spitzer} IRS spectrum of NGC 4258, extracted from spectral cubes in the SL+LL overlap region.
  The H$_2$ 0-0 S(L) pure rotational emission lines indicate a large mass of warm ($>240$ K) molecular gas.
  Ionized gas lines and PAH features are also indicated.
  (b) Spectral energy distribution, measured in IRS extraction region. Photometry was extracted from archival images:
  GALEX (NUV, FUV), SDSS (u, g, r, i, z), 2MASS (J, H, K), {\it Spitzer} IRAC (3.6, 4.5, 5.8, 8.0 $\mu$m), MIPS 24 $\mu$m,
  {\it Herschel} PACS (70, 100, 160 $\mu$m), and SPIRE (250, 350 $\mu$m). The SED was modeled using {\sc MAGPHYS} \citep{dce08}.}
\end{figure*}

\begin{figure*}[t]
\centerline{\includegraphics[width=0.8\linewidth]{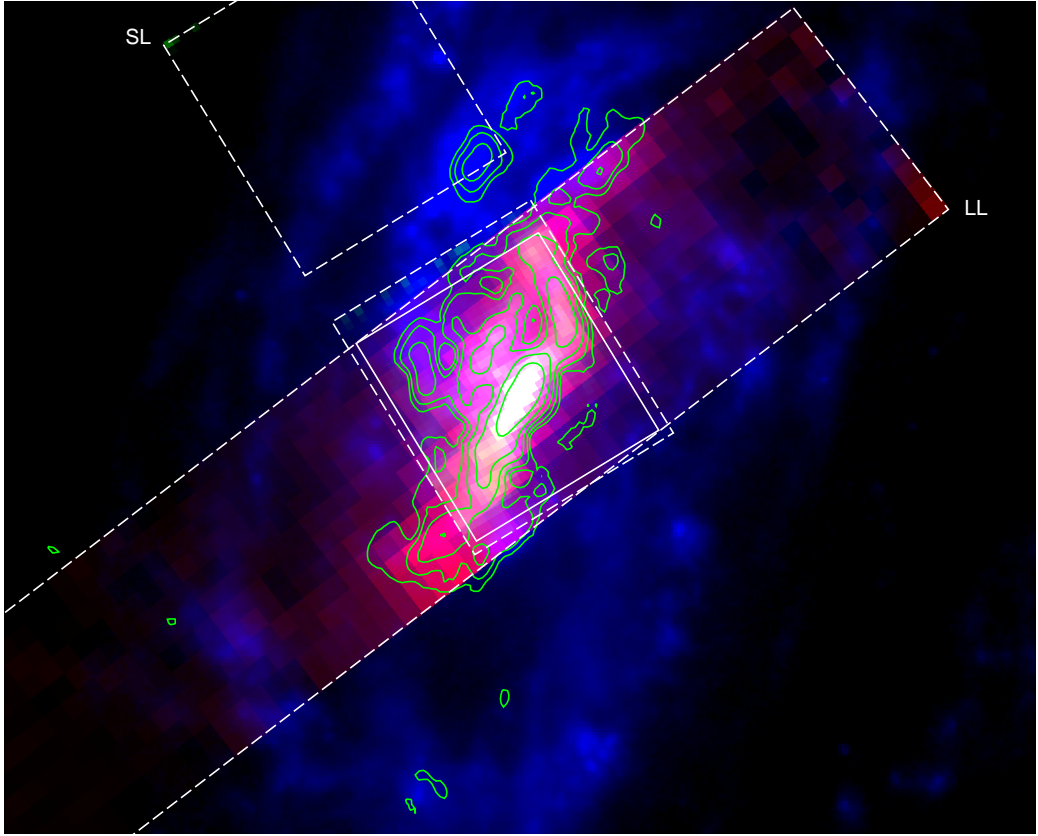}}
\caption{NGC 4258 central region. IRAC 8 $\mu$m image (blue) combined with H$_2$ 0-0 S(3) emission from the SL1 data cube (green), and  H$_2$ 0-0 S(1) emission from the
  LL2 data cube (red). The IRAC 8 $\mu$m band is dominated by hot dust and 7.7 $\mu$m PAH emission. Contours of the CO 1-0 map from BIMA (green) are
  overlaid \citep{htr03}. Note the close correspondence between CO 1-0 emission and H$_2$ emision, suggesting shock-enhancement of CO. The smaller SL and larger
  LL map footprints are indicated by dashed lines. The overlap region used to extract the IRS spectrum and multiwavelength SED is indicated by the central
  solid box ($50.7\arcsec \times 55.0 \arcsec$, or $1.76 \times 1.92$ kpc).}
\end{figure*}

\begin{figure*}[t]
\centerline{\includegraphics[width=0.7\linewidth]{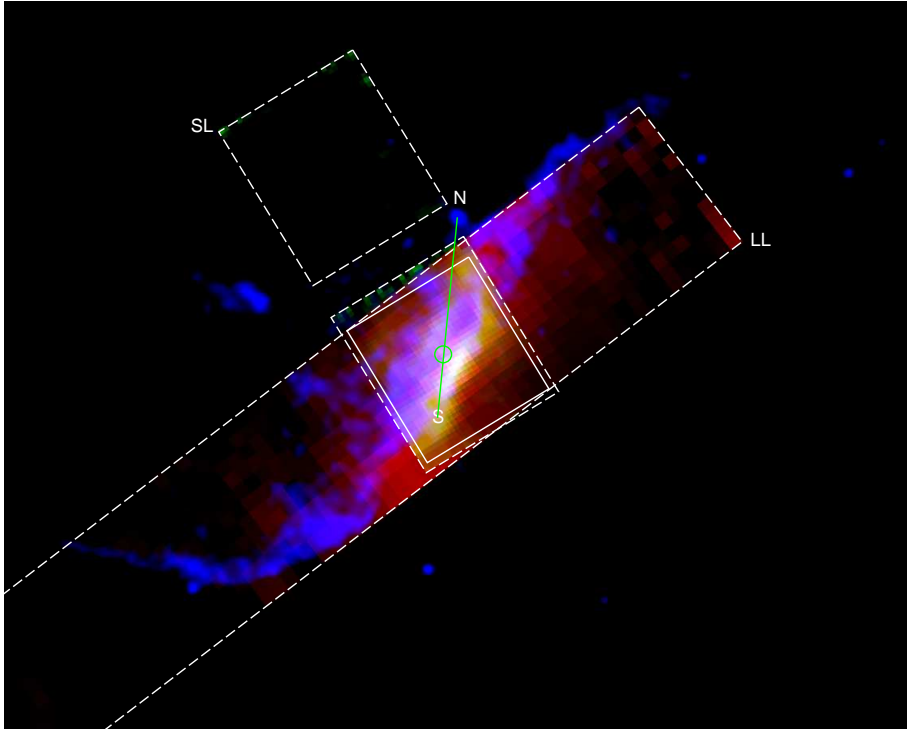}}
\caption{NGC 4258 anomalous arms. {\it Chandra} ACIS 0.3-8 keV image (blue, with point sources removed) combined with H$_2$ S(3) (green, SL1) and H$_2$ S(1) (red, LL2).
  The H$_2$ emission follows the X-ray emission from the anomalous arms, at least to where they bend out of the plane of the host galaxy. The bright, primary
  ridge of H$_2$ emission is displaced 6\arcsec~ SW of the AGN (green circle). The axis of the radio jet (green line) and radio jet hot spots (N, S) are shown for reference.
  Note that the radio jet is misaligned with the anomalous arms because it tilts 30\arcdeg out of the plane of the galaxy \citep{cdg00}.}
\end{figure*}

\begin{figure*}[t]
\centerline{\includegraphics[width=0.9\linewidth]{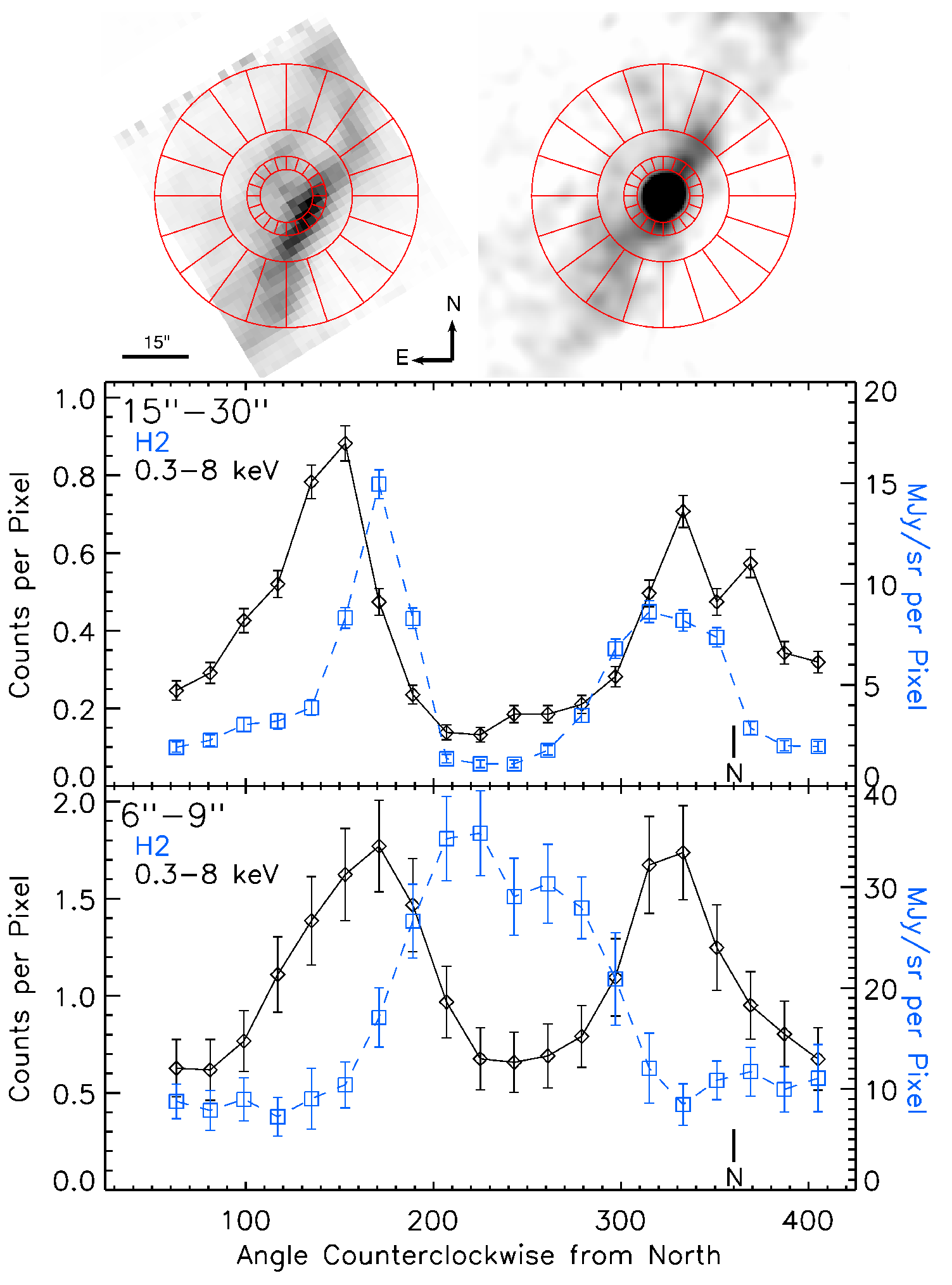}}
\caption{Comparison of {\it Spitzer} H$_2$ 0-0 S(3) and {\it Chandra} 0.3-8 keV X-ray azimuthal emission profiles for NGC 4258. The top two panels show the H$_2$ 0-0 S(3)
  flux (left) and background-subtracted X-ray counts (right). The bullseye pattern shows the regions used to extract the flux for the azimuthal
  profiles. The bottom two panels show that while the H$_2$ emission follows the PA of the X-ray emission at 15-30\arcsec, it is anticorrelated with the
  inner X-ray jets at 6-9\arcsec.}
\end{figure*}

\eject


\begin{thebibliography}{}
  
\bibitem[Appleton et al. (2013)]{a13} Appleton, P. N., Guillard, P., Boulanger, F. et al. 2013, ApJ, 777, 66
\bibitem[Appleton et al. (2006)]{a06} Appleton, P. N., Xu, K. C., Reach, W. et al. 2006, ApJ, 639, L51
\bibitem[Bolatto et al. (2013)]{bwl13} Bolatto, A. D., Wolfire, M., \& Leroy, A. K. 2013, ARA\&A, 51, 207
\bibitem[Cecil et al. (2000)]{cdg00} Cecil, G., Greenhill, L. J., DePree, C. G. et al. 2000, ApJ, 536, 675
\bibitem[Cecil, Morse, \& Veilleux (1995)]{cmv95} Cecil, G., Morse, J. A., \& Veilleux, S. 1995, ApJ, 452, 613
\bibitem[Cecil, Wilson, \& De Pree (1995)]{cwd95} Cecil, G., Wilson, A. S., \& De Pree, 1995, ApJ, 440, 181
\bibitem[Cluver et al. (2010)]{cab10} Cluver, M. E., Appleton, P. N., Boulanger, F. et al. 2010, ApJ, 710, 248
\bibitem[Courtes \& Cruvellier (1961)]{cc61} Courtes, \& Cruvellier, 1961, Compt. Rend. Acad. Sci. Paris, 253, 218
\bibitem[Cox \& Downes (1996)]{cd96} Cox, P. \& Downes, D. 1996, ApJ, 473, 219
\bibitem[Croton et al. (2006)]{csw06} Croton, D. J., Springel, V., White, S. D. M. et al. 2006, MNRAS, 365, 11
\bibitem[da Cunha et al. (2008)]{dce08} da Cunha, E., Charlot, S., \& Elbaz, D. 2008, MNRAS, 388, 1595
\bibitem[Emonts et al. (2005)]{emt05} Emonts, B. H. C., Morganti, R., Tadhunter, C. N. et al. MNRAS, 362, 931
\bibitem[Fabian et al. (2006)]{fst06} Fabian, A. C., Sanders, J. S., Taylor, G. B., et al. 2006, MNRAS 366, 417
\bibitem[Fabian et al. (2000)]{fse00} Fabian, A. C., Sanders, J. S., Ettori, S. et al. 2000, MNRAS, 318, 65
\bibitem[Greenhill et al. (1995)]{gjm95} Greenhill, L. J., Jiang, D. R., Moran, J. M., \& Reid, M. J. 1995, ApJ, 440, 619
\bibitem[Guillard et al. (2009)]{gbp09} Guillard, P., Boulanger, F., Pineau des Forets, G., \& Appleton, P. N. 2009, A\&A, 502, 515
\bibitem[Guillard et al. (2012)]{gbp12} Guillard, P., Boulanger, F., Pineau des Forets, G. et al. 2012, ApJ, 749, 158
\bibitem[Guillard et al. (2012)]{goe12} Guillard, P., Ogle, P. M., Emonts, B. H. C. et al. 2012, ApJ, 747, 95
\bibitem[Helfer et al. (2003)]{htr03} Helfer, T. T., Thornley, M. D., Regan, M. W. et al. 2003, ApJS, 145, 259
\bibitem[Ho et al. (1997)]{h97} Ho, L. C., Fillipenko, A. V., \& Sargent, W. L. W. 1997, ApJS, 112, 315
\bibitem[Kennicutt (1998)]{k98} Kennicutt, R. C. 1998, ApJ, 498, 541
\bibitem[Kraft et al. (2007)]{knb07} Kraft, R. P., Nulsen, P. E. J., Birkenshaw, M. et al. 2007, ApJ, 665, 1129
\bibitem[Krause, Fendt, \& Neininger (2007)]{kfn07} Krause, M., Fendt, C., \& Neininger, N. 2007, A\&A, 467, 1037
\bibitem[Laine et al. (2010)]{lkt10} Laine, S., Krause, M., Tabatabaei, F. S., \& Siopis, C. 2010, AJ, 140, 1084
\bibitem[Lebrun et al. (1983)]{l83} Lebrun, F., Bennett, K., Bignami, G. F. et al. 1983, ApJ, 274, 231
\bibitem[Massaro et al. (2009)]{mcg09} Massaro, F., Chiaberge, M., Grandi, P. et al. 2009, ApJ, 692, 123
\bibitem[Miyoshi et al. (1995)]{mmh95} Miyoshi, M., Moran, J., Hernstein, J. et al. 1995, Nature, 373, 127
\bibitem[Morganti et al. (2005)]{mto05} Morganti, R., Tadhunter, C. N., \& Osterloo, T. A. 2005, A\&A, 444, L9
\bibitem[Martin et al. (1989)]{mrn89} Martin, P., Roy, J. R., Noreau, L., \& Lo, K. Y. 1989, A\&A, 345, 707
\bibitem[Hardcastle et al. (2012)]{hmh12} Hardcastle, M. J., Massaro, F., Harris, D. E. et al. 2012, MNRAS, 424, 1774
\bibitem[Ogle et al. (2010)]{obg10} Ogle, P., Boulanger, F., Guillard, P. et al. 2010, ApJ, 724, 1193
\bibitem[Ogle et al. (2007)]{oaa07} Ogle, P., Antonucci, R., Appleton, P. N., \& Whysong, D. 2007, ApJ, 668, 699
\bibitem[Roussel et al. (2007)]{rhh07} Roussel, H., Helou, G., Hollenbach, D. J.  et al. 2007, ApJ, 669, 959
\bibitem[Roussel et al. (2001)]{rsv01} Roussel, H., Sauvage, M., Vigroux, L., \& Bosma, A., 2001, A\&A, 372, 427
\bibitem[Sandstrom et al. (2013)]{slw13} Sandstrom, K. M., Leroy, A. K., Walter, F. et al. 2013, ApJ, 777, 5
\bibitem[Savage \& Mathis (1979)]{sm79} Savage, B. D. \& Mathis, J. S. 1979, ARA\&A, 17, 73
\bibitem[Sutherland \& Bicknell (2007)]{sb07} Sutherland, R. S. \& Bicknell, G. V. 2007, ApJS, 173, 37
\bibitem[Smith et al. (2007a)]{sad07a} Smith, J. D. T., Armus, L., Dale, D. A. et al. 2007a, PASP, 119, 1133
\bibitem[Smith et al. (2007b)]{sdd07b} Smith, J. D. T., Draine, B. T., Dale, D. A. et al. 2007b, ApJ, 656, 770
\bibitem[van der Kruit, Oort, \& Mathewson (1972)]{vom72} van der Kuit, P.C., Oort, J. H., \& Mathewson, D. S. 1972, A\&A, 21, 169
\bibitem[Wagner \& Bicknell (2011)]{wb11} Wagner, A. Y., \& Bicknell, G. V. 2011, ApJ, 728, 29
\bibitem[Wilson et al. (2001)]{wyc01} Wilson, A. S., Yang, Y., \& Cecil, G. 2001, ApJ, 560, 689
\bibitem[Yang et al. (2007)]{ylw07}  Yang, Y., Li, B., Wilson, A. S., \& Reynolds, C. S. 2007, ApJ, 660, 1106 

\end{thebibliography}
\end{document}